\documentclass[reprint]{revtex4-1}
\usepackage{amsmath}
\usepackage{amsfonts}
\usepackage{mathrsfs}
\usepackage{gensymb}
\usepackage{bbm}
\usepackage{dsfont}

\usepackage{chemformula}

\definecolor{webgreen}{rgb}{0,.5,0}
\definecolor{webbrown}{rgb}{.6,0,0}
\definecolor{grigio}{rgb}{.85,.85,.85} 
\definecolor{RoyalBlue}{rgb}{0.0, 0.14, 0.4}
\definecolor{skyblue1}{rgb}{0.45,0.62,0.81}
\definecolor{skyblue2}{rgb}{0.2,0.39,0.64}
\definecolor{skyblue3}{rgb}{0.13,0.29,0.53}
\definecolor{scarlet1}{rgb}{0.93,0.16,0.16}
\definecolor{scarlet2}{rgb}{0.8,0,0}
\definecolor{scarlet3}{rgb}{0.64,0,0}

\usepackage{hyperref}
\hypersetup{%
    colorlinks=true, linktocpage=true, pdfstartpage=1, pdfstartview=FitV,%
    breaklinks=true, pdfpagemode=UseNone, pageanchor=true, pdfpagemode=UseOutlines,%
    plainpages=false, bookmarksnumbered, bookmarksopen=true, bookmarksopenlevel=1,%
    hypertexnames=true, pdfhighlight=/O,
    urlcolor=webbrown, linkcolor=RoyalBlue, citecolor=webgreen, 
    pdftitle={Chemical Cloaking},%
    pdfauthor={Francesco Avanzini},%
    pdfsubject={},%
    pdfkeywords={},%
    pdfcreator={pdfLaTeX},%
    pdfproducer={LaTeX REVTeX}%
}

\DeclareMathAlphabet{\mathpzc}{OT1}{pzc}{m}{it}

\newcommand{\cloaking}{i}
\newcommand{\gradient}{o}

\begin{document}

\title{Chemical Cloaking}
\newcommand\unilu{\affiliation{Complex Systems and Statistical Mechanics, Department of Physics and Material Sciences, University of Luxembourg, L-1511 Luxembourg}}
\author{Francesco Avanzini}
 \email{francesco.avanzini@uni.lu}
\unilu
\author{Gianmaria Falasco}
 \email{gianmaria.falasco@uni.lu}
\unilu
\author{Massimiliano Esposito}
 \email{massimiliano.esposito@uni.lu}
\unilu

\date{\today}

\begin{abstract}
Hiding an object in a chemical gradient requires to suppress the distortions it would naturally cause on it.
To do so, we propose a strategy based on coating the object with a chemical reaction-diffusion network which can act as an active cloaking device. 
By controlling the concentration of some species in its immediate surrounding, the chemical reactions redirect the gradient as if the object was not there. 
We also show that a substantial fraction of the energy required to cloak can be extracted from the chemical gradient itself. 
\end{abstract}

\maketitle


Inhomogeneous concentrations are a hallmark of out-of-equilibrium phenomena and not surprisingly play a fundamental role in biology. Common examples are patterns formation in morphogenesis~\cite{Kondo1616, Kretschmer2016}, chemical waves in signaling~\cite{Deneke2018}, and chemotaxis~\cite{Swaney2010, Levine2013} where single cells~\cite{Berg1977, Bialek2005, Endres2008} or cell clusters~\cite{Camley2017, Camley2018} sense gradients to detect energy sources or hazards. White blood cells for instance exploit chemical gradients to detect unidentified substances, generic wounds, and cancer cells~\cite{Eisenbach2004}. 
One strategy to avoid being detected is to avoid the distortions on the gradient that any generic object embedded into it would produce. In this letter, we propose an active strategy to leave the chemical gradient unchanged: by coating an object with a reaction-diffusion system, the chemical gradient around the object is restored as if the object was not there. 
Previous works on cloaking adopted a very different strategy based on coating the object with a metamaterial~\cite{Smith2004, Kadic2013} with transport properties tuned in such a way to directed the field around the object.  This was initially conceived to achieve optical invisibility~\cite{Pendry2006, Leonhardt2006} by tuning the permittivity and permeability of the material, and was also experimentally realized~\cite{Valentine2009, Gabrielli2009, Zhang2011, Ergin2011}. Similar strategies~\cite{Shalaev2008, Chen2010, Pendry2012} were later used to tune diffusion tensors and cloak either heat or mass flows~\cite{Guenneau2012, Guenneau2013, Florez2017, Guenneau2017}. Cloaking from quantum matter waves~\cite{Greenleaf2008, Zhang2008} or hydrodynamic environment was also considered~\cite{Park2019, Zou2019}. 
Crucially however, all these works share the same fundamental approach: they design the space variation of a tensor transport property (e.g., permittivity, permeability, diffusion) and thus correspond to passive mechanisms of cloaking. Our approach is instead based on an active process using chemical reactions which leaves the diffusion coefficients unchanged.
One may speculate about whether such strategy is or could be implemented in biosystems where chemical reactions are ubiquitously present. 

\begin{figure}[p]\centering
\includegraphics[width=1.\columnwidth]{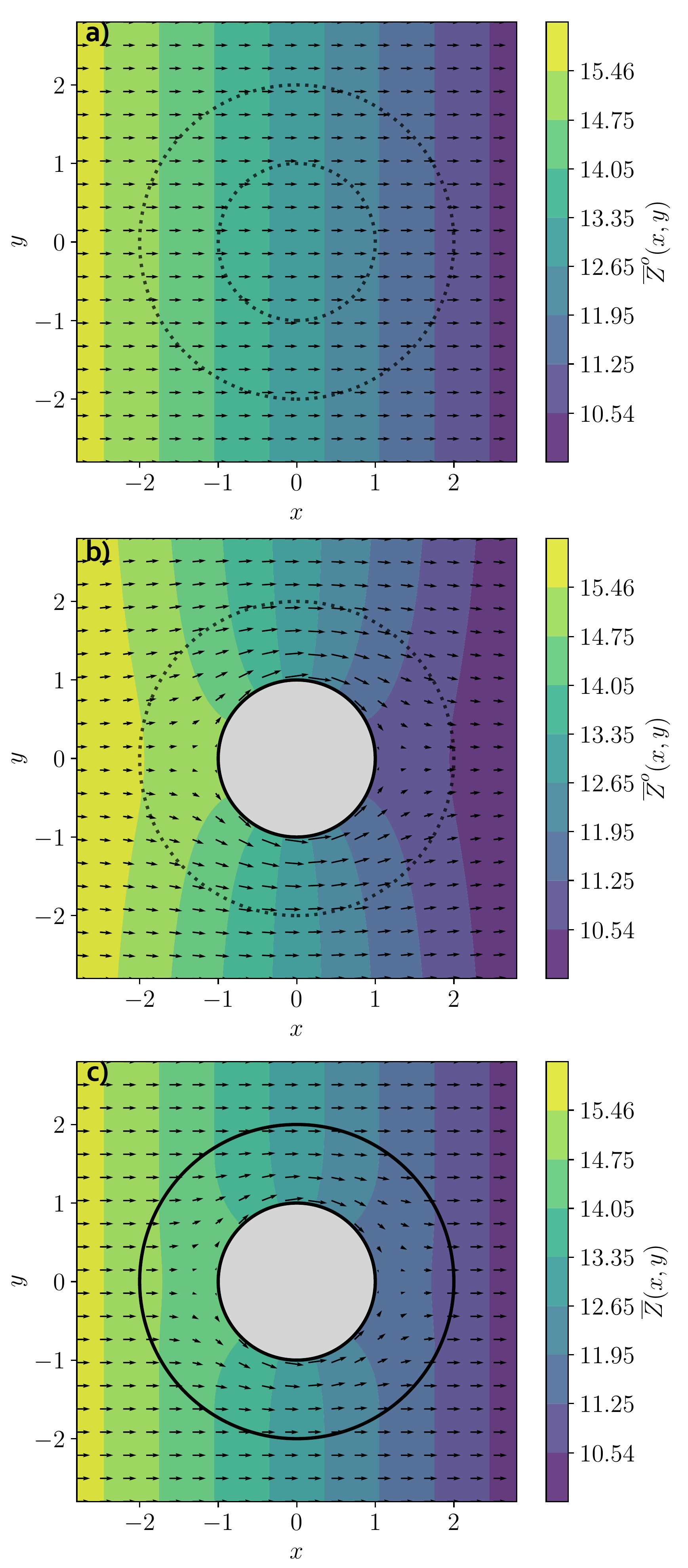}
\caption{Steady-state concentration and diffusion flow field in a vertical stripe domain of width $2 L$ maintained at two different concentrations $z_1$ and $z_2$ on each side a) in absence of an object b) in presence of that object of radius $R_1$ c) when the object is coated by a ring shaped cloaking device of width $R_2-R_1$. Here $L=2.8$, $z_1=15.8$ $z_2=10.2$, $D=1$, $R_1=1$ and $R_2=2$. We use $k_{+}/k_{-}$, $k_{-}/(k_{+})^2$ and $\sqrt{Dk_{-}}/k_{+}$ as units of measure for concentration, time and space, respectively.}
\label{fig_environment_plot}
\end{figure}

Lets consider a two-dimensional region of space $\Omega$ in which a species \ch{Z} freely diffuses, $\partial_t Z^{\gradient} = -\boldsymbol\nabla\cdot\boldsymbol{J}^{\gradient}_Z$ with the Fick's diffusion flux $\boldsymbol{J}^{\gradient}_Z=-D \boldsymbol \nabla Z^{\gradient}$, and $D$ the scalar (and constant) diffusion coefficient. At steady-state, the concentration profile is thus given by the Laplace equation, $\nabla^2 \overline{Z}^{\gradient}=0$, complemented by boundary conditions on the contour $\partial \Omega$. If an impermeable object is introduced in $\Omega$, the steady-state profile will be distorted. Our main finding in this letter is to design a ring shaped cloaking device which, when surrounding the object, can restore the outer steady-state profile to its pristine state $Z^{\gradient}$ (i.e. the one in absence of the object). The outer (inner) surface of the ring at radius $R_2$ ($R_1$) is permeable (impermeable) to \ch{Z}. The region between the outer and the inner surface 
 is denoted $\Omega_{\cloaking}$ and chemical reactions take place in it. The region inside the inner surface is the core that will embed the object. 
To enable analytical calculations, we will focus on a setup where $\Omega$ is a vertical stripe of width $2L$ centered around the origin with concentration $z_1$ on the left and $z_2$ on the right boundary. As illustrated in Fig.~\ref{fig_environment_plot}a, in its pristine state this setup gives rise to a linear steady-state profile 
\begin{equation}
\overline{Z}^{\gradient}(x,y)=\beta x +z_0,
\label{eq_linear_gradient}
\end{equation}
with $\beta=(z_2-z_1)/2L$ the slope of the gradient and $z_0=(z_2+z_1)/2$ the concentration at $x=0$. We exemplify in Fig.~\ref{fig_environment_plot}b how a circular object of radius $R_1$ distorts this profile. 

We now explain how our ring shaped cloaking device center around the origin can be used to restore any pristine concentration profile outside $R_2$ by enclosing an arbitrary object within $R_1$. 
To cloak and be compatible with matter conservation, the steady-state concentration of \ch{Z} inside the ring, $\overline{Z}^{\cloaking}$, must obey the following boundary conditions.
The internal concentration and its gradient must be continuous at $R_2$ with the outer pristine profile
\begin{align}
&\overline{Z}^{\cloaking}(R_2,\theta)=\overline{Z}^{\gradient}(R_2,\theta) \label{eq_bc_external_field},
\\
&\boldsymbol\nabla \overline{Z}^{\cloaking}(R_2,\theta)=\boldsymbol\nabla \overline{Z}^{\gradient}(R_2,\theta).
\label{eq_bc_external_grad}
\end{align}
Furthermore, the gradient must be tangent to the inner boundary at $R_1$ due to its impermeability 
\begin{equation}
\boldsymbol\nabla \overline{Z}^{\cloaking}(R_1,\theta)\cdot\hat{\boldsymbol r}=0.
\label{eq_bc_internal}
\end{equation}
We introduced polar coordinates~$(r,\theta)$ and the radial unit vector $\hat{\boldsymbol r}$. 
The boundary conditions, Eqs.~\eqref{eq_bc_external_field},~\eqref{eq_bc_external_grad} and~\eqref{eq_bc_internal} constitute the cloaking conditions. 
In Appendix~A, we show that a possible steady-state profile inside the cloaking device able to restore an arbitrary pristine profile $\overline{Z}^{\gradient}$ is
\begin{equation}
\small
\overline{Z}^{\cloaking}(r,\theta) =\frac{\partial_r\overline{Z}^{\gradient}(R_2,\theta)}{\mathrm d_r\mathcal R(R_2)}\left[\mathcal R(r)-\mathcal R(R_2)\right]+\overline{Z}^{\gradient}(R_2,\theta),
\label{eq_generalss} 
\end{equation}
with $\mathcal R(r)=(r-R_1)^2$.
In the case of the vertical stripe model producing the pristine profile \eqref{eq_linear_gradient}, $\overline{Z}^{\cloaking}$ is shown in Fig.~\ref{fig_environment_plot}c.

Since our strategy makes use of chemical reactions, $\overline{Z}^{\cloaking}(r,\theta)$ must now be obtained as a steady-state solution of the reaction-diffusion equation
\begin{equation}
\partial_t Z^{\cloaking}(x,y; t)=j^{\cloaking}(x,y; t)-\boldsymbol\nabla \cdot\boldsymbol J^{\cloaking}_Z(x,y; t),
\label{eq_rdequation}
\end{equation}
where $j^{\cloaking}$ is the net current of \ch{Z} produced by all the chemical reactions according to the mass-action kinetics~\cite{Groot1984, Pekar2005} and $\boldsymbol J^{\cloaking}_Z$ is the Fick's diffusion flux $\boldsymbol J^{\cloaking}_Z= -D\boldsymbol\nabla Z_{ \scriptscriptstyle{S}}$. We assumed that the diffusion coefficient inside the ring is identical to the one outside, $D$. 
We therefore need to specify the net reactions that at steady-state will give rise to $\overline j^{\cloaking}=\boldsymbol\nabla\cdot \overline{\boldsymbol J}^{\cloaking}_Z= -D\nabla^2 \overline{Z}^{\cloaking}$, where $\overline{Z}^{\cloaking}$ is our cloaking profile given in \eqref{eq_generalss}.
In general this problem has multiple solutions, as multiple sets of chemical reactions may lead to the same steady-state profile. 

\begin{figure}[t]\centering
\includegraphics[width=1.\columnwidth]{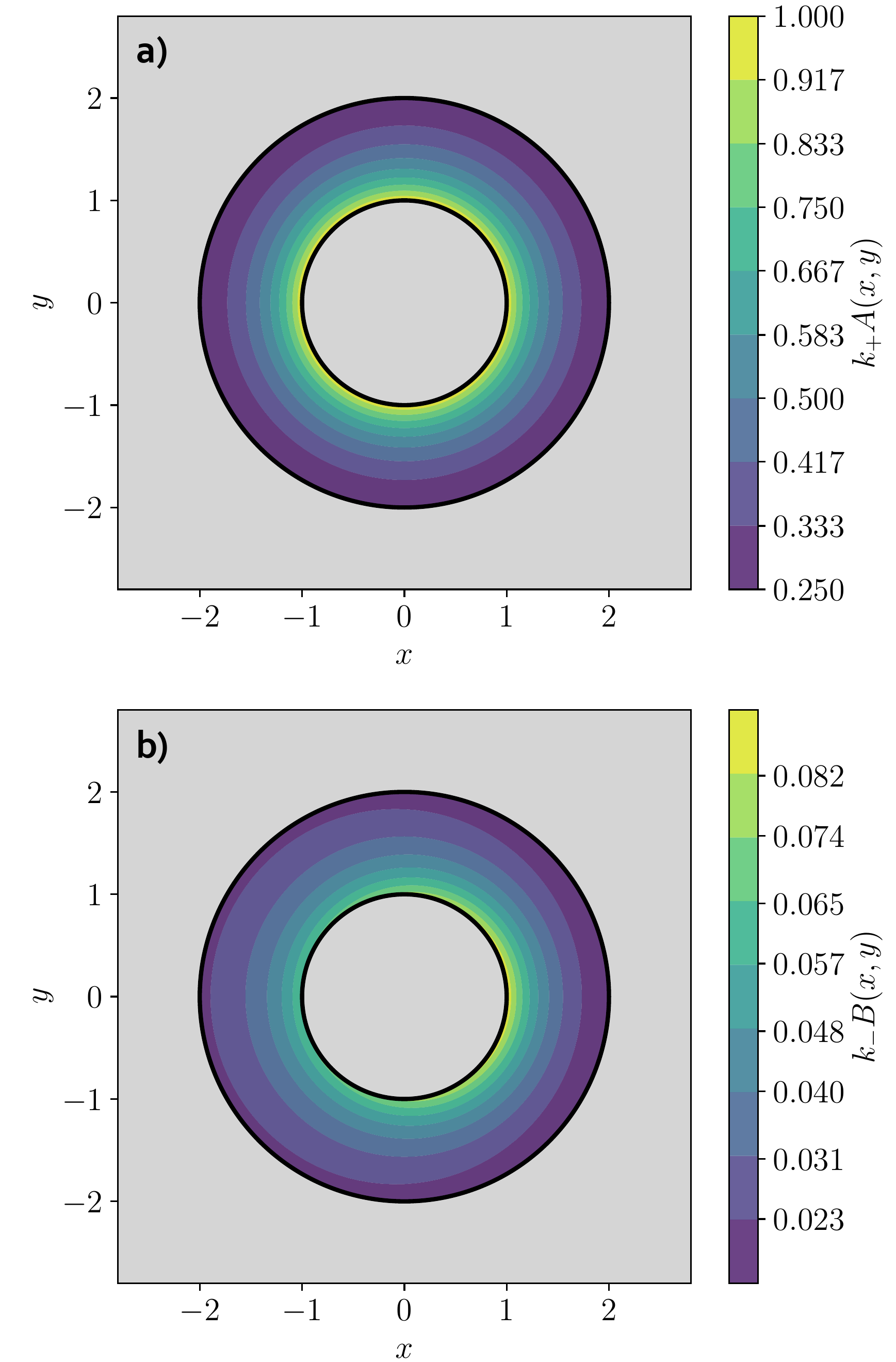}
\caption{Chemostatted species concentrations, a) $k_{+}A(x,y)$ and b) $k_{-}B(x,y)$, corresponding to Fig.~\ref{fig_environment_plot}c.}
\label{fig_concab_plot}
\end{figure}

To proceed we will now focus on our vertical stripe model.
Using \eqref{eq_linear_gradient} in \eqref{eq_generalss}, the steady-state concentration becomes
\begin{equation}
\overline{Z}^{\cloaking}(r,\theta)=\frac{\beta\cos\theta}{2}\left\{\frac{(r-R_1)^2}{R_2-R_1}+R_2+R_1\right\}+ z_0.
\label{eq_specificss}
\end{equation} 
After a simple manipulation, we find that the net steady-state reaction current, $\overline{j}^{\cloaking}=-D\nabla^2\overline{Z}^{\cloaking}$, can be expressed as
\begin{equation}
\overline{j}^{\cloaking}(r,\theta)=k_{+}A(r)\overline{Z}^{\cloaking}(r,\theta)-k_{-}B(r,\theta)(\overline{Z}^{\cloaking}(r,\theta))^2,
\label{eq_ssrctcurrent}
\end{equation}
where
\begin{align}
&k_{+}A(r)=\frac{D}{r^2},
\label{eq_achem}
\\
&k_{-}B(r,\theta)=D\frac{\beta r(2r-R_1)\cos\theta+z_0(R_2-R_1)}{r^2(R_2-R_1)\left(\overline{Z}^{\cloaking}(r,\theta)\right)^2}.
\label{eq_bchem}
\end{align}
This current can be interpreted as the mass-action reaction current produced by the reaction 
\begin{equation}
\ch{A + Z <=>[ $k_{+}$ ][ $k_{-}$ ] B + 2 Z} ,
\label{eq_crn}
\end{equation}
provided that $B(r,\theta)$ is non-negative in $\Omega_{\cloaking}$. 
Indeed, in this case both $A$ and $B$ can be interpreted as imposed concentrations which do not enter the dynamics, i.e. as chemostatted species~\cite{Falasco2018a}. This means that external mechanisms must actively maintain these concentrations and that the cloaking device must be fueled to operate.   
The non-negativity condition on $B$ holds if 
\begin{equation}
z_0\geq \frac{\left|\beta\right|R_2(2R_2-R_1)}{R_2-R_1}.
\label{eq_constraintaverage2}
\end{equation}
This constraint connects the properties of the pristine gradient, $z_0$ and $\beta$, to the geometry of the cloaking device, $R_1$ and $R_2$.
However, as discussed in Appendix~B, a linear stability analysis of the reaction-diffusion equation in $\Omega$ shows that the cloaking solution is stable only if the stronger constraint 
\begin{equation}
z_0\geq \frac{\left|\beta\right|R_2(3R_2-R_1)}{R_2-R_1} \label{eq_threshold_z0}
\end{equation}
is satisfied. 
The chemostatted concentrations \eqref{eq_achem} and \eqref{eq_bchem} corresponding to the vertical strip model used for Fig.~\ref{fig_environment_plot} are plotted in Fig.~\ref{fig_concab_plot}.
One observes that the angular dependence of $B(r,\theta)$ is weak. Indeed, for large values of $ z_0$, $B(r,\theta)\simeq{D}/{z_0 r^2}$.
We thus designed a ring shaped cloaking device containing the simple reaction \eqref{eq_crn} kept out-of-equilibrium by the chemostatted species \eqref{eq_achem} and \eqref{eq_bchem}. This device would make any object in its core undetectable via concentration profile measurements by preserving the pristine concentrations profile \eqref{eq_linear_gradient}.


We will now assess the energetic cost needed to maintain the cloaking state. 
While part of the energy is provided by the chemostats, we will see that another part can be extracted from the chemical gradient. 
The entropy production rate in an isothermal system quantifies the energy dissipation. 
For a reaction-diffusion system at steady-state, it is given locally by~\cite{Avanzini2019a} 
\begin{equation}
T\overline{\dot{\sigma}}=\overline{\dot{w}}_{\text{chem}}-\boldsymbol\nabla \cdot \sum_{\alpha} \left( \overline{\mu}_\alpha \overline{\boldsymbol J}_\alpha \right) \geq 0,
\label{eq_balance_local_ss}
\end{equation}
where $\alpha$ labels the different species (for our model $\alpha=A,B,Z$), $\overline{\boldsymbol J}_\alpha$ their Fick fluxes, and $\overline{\mu}_\alpha= \mu_\alpha^{\theta} + RT \ln\overline{ [\alpha]}$ their chemical potentials. Here, $R$ is the gas constant, $T$ is the temperature of the solvent, $\mu_\alpha^{\theta}$ is the standard chemical potentials and $\overline{[\alpha]}$ is the steady-state species concentrations. 
The chemical work rate $\overline{\dot{w}}_{\text{chem}}$ is given by 
\begin{equation}
\overline{\dot{w}}_{\text{chem}}=\sum_{\alpha \in \text{chemstat}} \overline{\mu}_\alpha \overline{I}_\alpha
\end{equation}
and represents the chemical energy entering the system from the chemostats via the currents $\overline{I}_\alpha$. 
For our model $\overline{I}_A= \overline{j}^{\cloaking}-D_{A}\nabla^2A$ and $\overline{I}_B= -\overline{j}^{\cloaking}-D_{B}\nabla^2B$.

\begin{figure}[t]\centering
\includegraphics[width=1.\columnwidth]{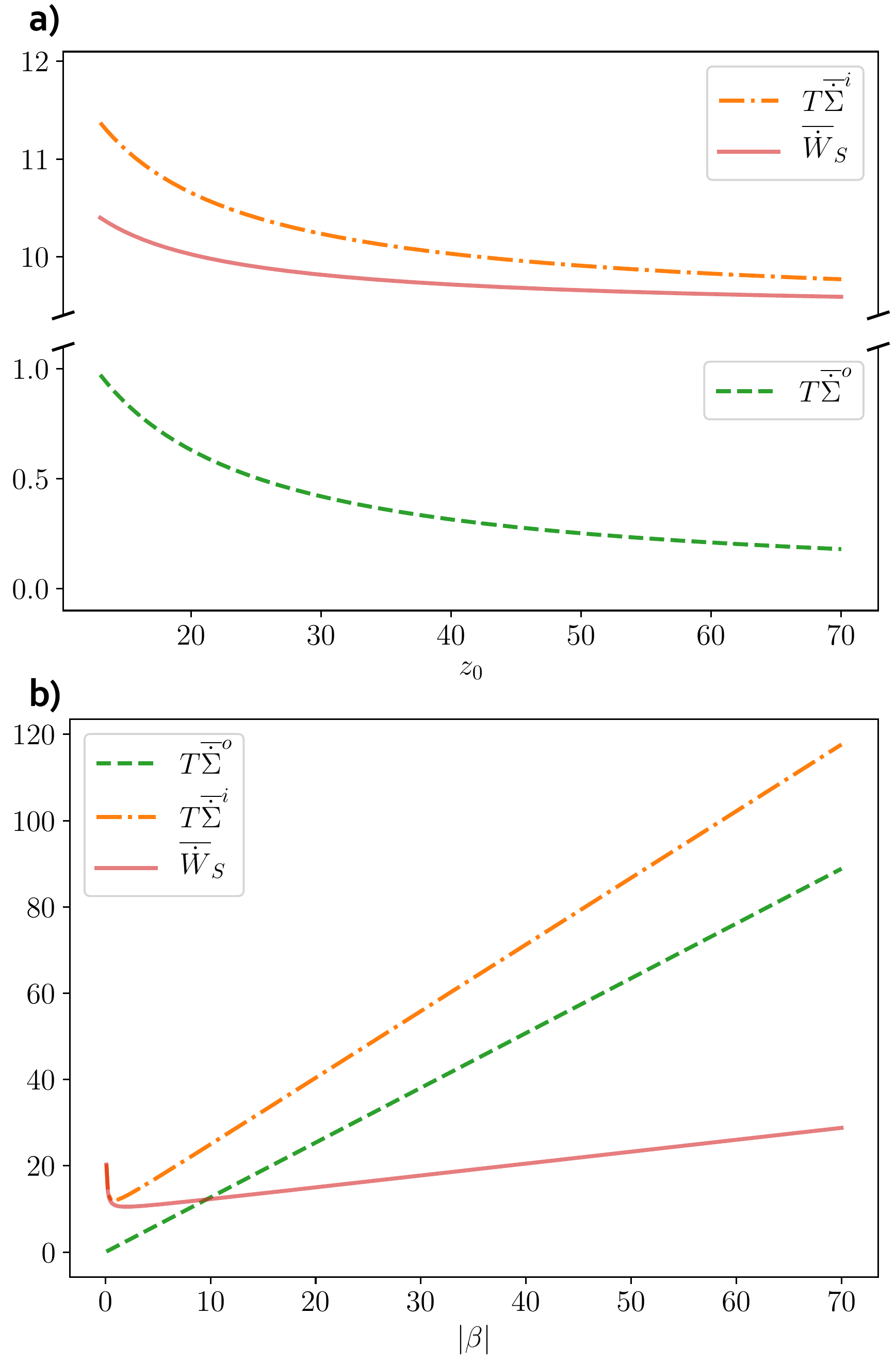}
\caption{Thermodynamic quantities as a function of a) $z_0$ (with $\beta=-1$) and b) $|\beta|$, corresponding to Fig.~\ref{fig_environment_plot}c. Here $k_{+}=k_{-}=1$, $D_A=D_B=1$ and we use $RT$ as units of measure for the energy.}
\label{fig_thermo_plot}
\end{figure}

Let us first introduce the dissipation for the pristine profile occurring in the region $\Omega_C$ that will be later occupied by the cloaking device (ring and core). Since it is caused purely by the diffusion of \ch{Z} in $\Omega_C$, integrating \eqref{eq_balance_local_ss} with $\overline{\dot{w}}_{\text{chem}}=0$ over $\Omega_C$ and using Gauss's theorem, it is given by 
\begin{equation}
T\overline{\dot{\Sigma}}^{\gradient}
=-\int_{\partial\Omega_C}\mathrm d\boldsymbol a \cdot(\overline{\mu}^{\gradient}_Z\overline{\boldsymbol J}^{\gradient}_Z) \geq 0.
\label{eq_DefDissPRistProf}
\end{equation} 
It can be interpreted as the energy entering the region $\Omega_C$ thought its boundary.
Turning back to the the cloaking device its dissipation is obtained by integrating \eqref{eq_balance_local_ss} over $\Omega_{\cloaking}$ and can be written as 
\begin{equation}
T\overline{\dot{\Sigma}}^{\cloaking}= \overline{\dot{W}}^{\cloaking}+ T\overline{\dot{\Sigma}}^{\gradient} \geq 0.
\label{eq_dissipation_rdsBis}
\end{equation}
The total work rate $\overline{\dot{W}}^{\cloaking}$ performed by the chemostats is given by
\begin{equation}
\overline{\dot{W}}^{\cloaking}=\overline{\dot{W}}_{\text{chem}}-\int_{\partial\Omega_{\cloaking}}\mathrm d \boldsymbol a\cdot\sum_{\alpha=A,B}\overline{\mu}_{\alpha}\overline{\boldsymbol J}_{\alpha}.
\label{eq_dissipation_rds}
\end{equation}
The first term comes from the direct integration of $\overline{\dot{w}}_{\text{chem}}$ over $\Omega_{\cloaking}$ and represents the chemical energy provided by the chemostats inside the cloaking device. 
The second term is obtained using Gauss's theorem from the diffusion part of~\ch{A} and~\ch{B} in \eqref{eq_balance_local_ss}. 
It represents the energy provided by the chemostats to maintain an heterogeneous concentration of \ch{A} and \ch{B} on the inner and outer boundary $\partial \Omega_{\cloaking}$ of the device against diffusion. 
The second contribution in~\eqref{eq_dissipation_rdsBis} is obtained using Gauss's theorem from the diffusion of \ch{Z} inside $\Omega_{\cloaking}$ and is thus given by $-\int_{\partial\Omega_{\cloaking}}\mathrm d\boldsymbol a \cdot(\overline{\mu}^{\cloaking}_Z\overline{\boldsymbol J}^{\cloaking}_Z)$ which using the cloaking condition~\eqref{eq_bc_external_field},~\eqref{eq_bc_external_grad} and~\eqref{eq_bc_internal} has the transparent interpretation of being the chemical energy provided by the pristine profile through the boundaries of the cloaking device \eqref{eq_DefDissPRistProf}.
Equation \eqref{eq_dissipation_rdsBis} is a key result of this letter. It shows that the dissipation of the cloaking device can be expressed as the energy provided by the chemostats on the one end and by the external pristine profile on the other end. While the latter is always non-negative, the former has no predefinite sign. When negative, the cloaking device is fully propelled by the energy provided by the external gradient. 
However, we did not find such a regime. We now reproduce the typical dissipative contributions observed by considering a given cloaking device (defined by the parameters $R_1$, $R_2$, $D_A$ and $D_B$) as a function of the features of the pristine profile ($z_0$ and $\beta$). As a function of $z_0$, Fig.~\ref{fig_thermo_plot}a shows that for a given slope of the gradient the total work rate to operate the cloaking device is a significant fraction of the total dissipation. Furthermore as $z_0$ increases, the energy provided by the gradient, $T\overline{\dot{\Sigma}}^{\gradient}$, tends to zero and the energy to operate the device is fully provided by the chemostats. However, in Fig.~\ref{fig_thermo_plot}b, when increasing $\beta$ and choosing the corresponding minimum $z_0$ (according to~\eqref{eq_threshold_z0}), the energy is predominantly provided by the gradient of pristine profile. In that limit the energy provided by the chemostats to cloak reaches approximatively $20 \%$ of the total dissipation.

Our work provides the first proof of principle that chemical cloaking is possible. It also raises interesting questions. Is there a general procedure to identify the chemical reactions cloaking in a generic gradient? Is there an optimal shape for the cloaking device which could be used to design self powered devices? How fast the cloaking conditions can be established? Are there chemical processes  in biosystems (e.g. self quorum quenching in bacteria colonies~\cite{Hense2015}) acting as cloaking devices? We leave the questions to future investigations.

We acknowledge funding from the European Research Council project NanoThermo (ERC-2015-CoG Agreement No.~681456). 
 


\paragraph*{Appendix A -} We describe the procedure to find the steady-state profile satisfying the cloaking conditions~\eqref{eq_bc_external_field},~\eqref{eq_bc_external_grad} and~\eqref{eq_bc_internal}. 
One first notes that the cloaking conditions impose three and not four constraints because the equality at $r=R_2$ between the concentrations $\overline{Z}^{\cloaking}(R_2,\theta)=\overline{Z}^{\gradient}(R_2,\theta)$ (Eq.~\eqref{eq_bc_external_field}) ensures the equality between the tangential component of the concentration gradients at $r=R_2$, $\partial_\theta \overline{Z}^{\cloaking}(R_2,\theta)=\partial_\theta \overline{Z}^{\gradient}(R_2,\theta)$ (Eq.~\eqref{eq_bc_external_grad}). 
Motivated by this, we make the following ansatz 
\begin{equation}
\overline{Z}^{\cloaking}(r,\theta)=\zeta(\theta)\mathcal R(r)+\Gamma(\theta)\label{app_eq_general_expression_ss},
\end{equation}
with a radial function, $\mathcal R(r)$, and two angular functions, $\zeta(\theta)$ and $\Gamma(\theta)$ to be determined by the three cloaking conditions. 
Using the radial component $\hat{\boldsymbol r}$ of Eq.~\eqref{eq_bc_external_grad}, we find
\begin{equation}
\zeta(\theta) = \frac{\partial_r \overline{Z}^{\gradient}(R_2,\theta)}{\mathrm d_r\mathcal R (R_2)},\label{app_eq_zeta}
\end{equation}
while Eq.~\eqref{eq_bc_external_field} specifies $\Gamma(\theta)$ as 
\begin{equation}
\Gamma(\theta)= \overline{Z}^{\gradient}(R_2,\theta) - \zeta(\theta)\mathcal R(R_2).\label{app_eq_gamma}
\end{equation}
We now impose $\mathrm d_r\mathcal R(R_1)=0$ to ensure that the constraint~\eqref{eq_bc_internal} is satisfied. 
A possible solution that we will use is $\mathcal R(r)=(r-R_1)^2$, but other choices could be $ \mathcal R(r)=(r-R_1)^n$ with $n\in\mathbb N_{>2}$ or $ \mathcal R(r) = r + R^2_1/r$. 
This last result combined with the substituting of~\eqref{app_eq_zeta} and~\eqref{app_eq_gamma} in Eq.~\eqref{app_eq_general_expression_ss} finally gives the steady-state concentration~\eqref{eq_generalss}. 


\paragraph*{Appendix B -} We perform linear stability analysis by perturbing the concentration $Z(r,\theta; t)$ around the steady-state solution~\eqref{eq_specificss} inside the cloaking device, $\Omega_{\cloaking}$, and the pristine profile \eqref{eq_linear_gradient} outside it, $\Omega_{\gradient}$, overall denoted by $\overline{Z}$. 
Extending the reaction-diffusion dynamics \eqref{eq_rdequation} to the outside of the cloaking device by assuming that
\begin{equation}
\small
j^{\cloaking}(r,\theta;t)=k_{+}A(r)Z^{\cloaking}(r,\theta; t)-k_{-}B(r,\theta)(Z^{\cloaking}(r,\theta;t))^2,
\label{eq_rctcurrent}
\end{equation}
is vanishing outside the device, and linearizing for the perturbation $Z^{\text{p}}(r,\theta;t)=Z(r,\theta;t)-\overline{Z}(r,\theta)$, we get
\begin{equation}
\partial_tZ^{\text{p}}(r,\theta;t)=- \underbrace{(-D\nabla^2+V(r,\theta))}_{=:\hat{\mathbb L}}Z^{\text{p}}(r,\theta;t),
\end{equation}
where $V(r,\theta)=2k_{-}B(r,\theta)\overline{Z}^{\cloaking}(r,\theta)-k_{+}A(r)$ inside the device and vanishes outside.
The perturbation vanishes at the boundaries by construction and is assumed to be square-integrable.
The operator $\hat{\mathbb L}$ is hermitian, and its eigenvalues are thus real.
Stability can be proven if all eigenvalues are also positive to ensure the decay of the perturbation.  
A sufficient condition for this to happen is that $V(r,\theta)$ is always positive because in this case
\begin{equation}
\int_{\Omega_{\cloaking}\cup\Omega_{\gradient}}\mathrm dx\mathrm dy\text{ }\psi(r,\theta)\hat{\mathbb L}\psi(r,\theta) \geq 0
\end{equation}
for every square-integrable function $\psi$. A positive $V(r,\theta)$ is granted when the condition \eqref{eq_threshold_z0} is satisfied.

\bibliography{biblio}

\end{document}